\documentclass[12pt]{article}

\usepackage{amsmath,amssymb,amsfonts,amsbsy}
\usepackage{cite}
\usepackage{graphicx}
\usepackage{wrapfig}
\usepackage[font=small,labelfont=bf,labelsep=period]{caption}
\usepackage{bm}

\begin{document}
\addcontentsline{toc}{subsection}{{CONSTITUENT QUARK REST ENERGY
 AND WAVE FUNCTION ACROSS THE LIGHT CONE}\\
{\it M.V.~Bondarenco}}

\setcounter{section}{0}
\setcounter{subsection}{0}
\setcounter{equation}{0}
\setcounter{figure}{0}
\setcounter{footnote}{0}
\setcounter{table}{0}

\begin{center}
\textbf{CONSTITUENT QUARK REST ENERGY\\
 AND WAVE FUNCTION ACROSS THE LIGHT CONE}

\vspace{5mm}

M.V.~Bondarenco

\vspace{5mm}

\begin{small}
  \emph{Kharkov Institute of Physics and Technology, 1 Academicheskaya St.,
61108 Kharkov, Ukraine} \\
  \emph{E-mail: bon@kipt.kharkov.ua}
\end{small}
\end{center}

\vspace{0.0mm} 

\begin{abstract}
  It is shown that for a constituent quark in the intra-nucleon
  self-consistent field the spin-orbit interaction lowers the quark
  rest energy to values $\sim100$ MeV, which agrees with the DIS
  momentum sum rule. The possibility of violation of the spectral
  condition for the light-cone momentum component of a bound quark is discussed.
\end{abstract}

\vspace{7.2mm}

Given the information from the DIS momentum sum rule \cite{TW} that
half of the nucleon mass is carried by gluons, one may lower the
expectations for the constituent quark mass from $M_N/3$ to
$\lesssim M_N/6$. This may also provide grounds for formation of a
gluonic mean-field in the nucleon, permitting single-quark
description of the wave-function, and (by analogy with the situation
in atomic mean fields), making LS-interaction dominant over the SS.
\footnote{Neglecting the SS-interaction and employing the
single-quark model for the nucleon wave function, one must be ready
that the accuracy is not better than
$\frac{M_\Delta-M_N}{M_N}\sim30\div40\%.$} Here, adopting the
mentioned assumptions for the nucleon wave function, we will show
that the constituent quark mass $\sim 50\div100$ MeV, in fact,
results from the empirical values for the nucleon mean charge radius
and the magnetic moment. We will also address the issue of valence
quark wave function continuity and sizeable value at zero
light-front momentum, which owes to rather low a value of the
constituent quark mass, and discuss phenomenological implications
thereof.

\paragraph{Wave function ansatz}

For a spherically symmetric ground state, by parity reasons, the
upper Dirac component has orbital momentum $l=0$, whereas the lower
one has $l=1$ and flipped spin. This is accounted for by writing the
single-quark wave function
\begin{equation}\label{psi-q}
    \psi_q(\mathbf{r},t)=e^{-i\kappa^0t}\left(
                                        \begin{array}{c}
                                          \varphi(r)w \\
                                          -i\bm{\sigma}\cdot\frac{\mathbf{r}}r \chi(r)w \\
                                        \end{array}
                                      \right)\qquad \left(\kappa^0>0\right)
\end{equation}
with $w$ being an $\mathbf{r}$-independent Pauli spinor.

The relation between $\varphi$ and $\chi$ to a first rough
approximation is determined by the quark rest energy. For instance,
for a Dirac particle of mass $m$ moving in a static potential $V(r)$
(being either a 4th component of vector, or a scalar -- cf.
\cite{Critchfield}), in the ground state
\begin{equation}\label{g-f'}
\chi(r)=\frac1{m+E-V(r)}\varphi'(r).
\end{equation}
Thereat, a reasonable model may result from replacing the
denominator of (\ref{g-f'}) by its average
$\left\langle\kappa^0\right\rangle$ (which may somewhat differ from
$\kappa^0$ in (\ref{psi-q})):
\[
\chi(r)=\frac1{2\left\langle\kappa^0\right\rangle}\varphi'(r).
\]
For bag models \cite{TW}, the proportionality between $\chi$ and
$\varphi'$ holds exactly, but the particular shape of bag model wave
functions, especially $\chi$, may be not phenomenologically reliable
due to the influence of the bag sharp boundary. Instead, we prefer
to take $\varphi$ a smooth function for all $r$ from $0$ to
$\infty$:
\begin{equation}\label{psi-Dirac}
\psi_q(\bm\kappa)=\left(1+\frac1{2\left\langle\kappa^0\right\rangle}\gamma^0\bm{\gamma}\cdot\bm\kappa\right)\varphi(\kappa)\left(
                                                                                 \begin{array}{c}
                                                                                   w \\
                                                                                   0
                                                                                   \end{array}
                                                                               \right)
\equiv \varphi(\kappa)\left(
                                       \begin{array}{c}
                                         w \\
                                         \frac{\bm\sigma\cdot\bm\kappa}{2\left\langle\kappa^0\right\rangle}w
                                       \end{array}
                                     \right),
\end{equation}
\begin{equation}\label{ansatz}
\varphi(\kappa)\propto e^{-a^2\bm{\kappa}^2/2}.
\end{equation}
Thereby, the model involves only 2 parameters:
$\left\langle\kappa^0\right\rangle$ and the Gaussian radius $a$. It
may describe the valence quark component in the nucleon, and is
suitable for calculation of matrix elements of vector (conserved)
currents, whereas axial vector currents require the account of sea
quarks, which is beyond our scope here.
\begin{wrapfigure}[11]{R}{60mm}
  \centering 
  \vspace*{-8mm} 
  \includegraphics[width=57mm]{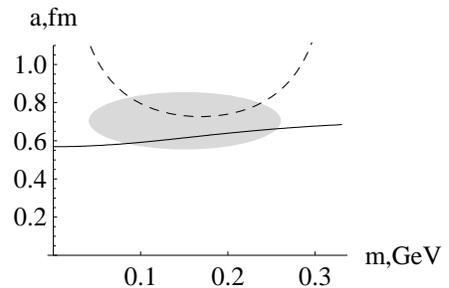}
  \caption{ Curves defined by Eq.~(\ref{rch-def}) (solid), Eq.~(\ref{mu-p}) (dashed) and the region of their consistency.}
  \label{fig:m-a}
\end{wrapfigure}
\paragraph{Proton mean charge radius and magnetic moment}
With recoilless spectators, our model can describe only formfactors
at $Q\ll M_N$. Equaing $Q/M_N$ to our accuracy $30\div40\%$, one
finds $Q<$0.4 GeV. In this region the decrease of the form-factors
is approximately linear \cite{TW}, whereby they are characterized
basically by the proton's magnetic moment and the mean charge (or
magnetic) radius.\footnote{By the isospin symmetry, neutron's
magnetic moment $\mu_n=-\frac23\mu_p$, and its magnetic radius is
about the same as proton's \cite{TW}.}

The form-factors are defined as overlaps of $\psi_q$ with $\psi'_q$,
of the same ground state, but with a shifted momentum and a
different polarization $w'$:
\begin{equation}\label{F1}
    \int\frac{d^3\kappa}{(2\pi)^3}\bar\psi'_q\left(\bm{\kappa}+\frac{\mathbf{Q}}2\right)\gamma^0\psi_q\left(\bm{\kappa}-\frac{\mathbf{Q}}2\right)=F_1(\mathrm{Q}^2)w'^+w,
    \qquad\left(F_1(0)=1\right)
\end{equation}
\begin{equation}\label{F2}
    \int\frac{d^3\kappa}{(2\pi)^3}\bar\psi'_q\left(\bm{\kappa}+\frac{\mathbf{Q}}2\right)\bm\gamma\psi_q\left(\bm{\kappa}-\frac{\mathbf{Q}}2\right)=F_2(\mathrm{Q}^2)w'^+\frac{i\left[\mathbf{Q}\times\bm\sigma\right]}{2\left\langle\kappa^0\right\rangle}w
\end{equation}
(the symmetric shift of momenta preserves the parity of the
integrand with respect to the origin, simplifying the calculations).
Since we neglect the recoil, $F_1$, $F_2$ may be associated equally
well with Dirac or with Sachs form-factors. That will be
non-contradictory provided in the exact identities relating those
types of form-factors
\begin{eqnarray}
  F_e = F_1-\frac{\mathbf{Q}^2}{4M_N^2}F_2, \qquad
  F_m = F_2+F_1
\end{eqnarray}
the last terms are small. For the proton,
$F_1(\mathbf{Q}^2)/F_2(\mathbf{Q}^2)\approx F_1(0)/F_2(0)=1/2.79$,
and within our accuracy $30\div40\%$ this contribution is indeed
negligible. Also, the Foldy term for the proton's mean square charge
radius $\frac{3F_2(0)}{2M_N^2}=\left(0.4 \mathrm{fm}\,\right)^2$ is
small compared to $\left\langle
r^2_{\mathrm{ch}}\right\rangle\approx1\,\mathrm{fm}^2$. Now with
Eqs.~(\ref{F1}, \ref{psi-Dirac}), let us actually evaluate the mean
square charge radius:
\begin{eqnarray}\label{rch-def}
    \left\langle
r^2_{\mathrm{ch}}\right\rangle&=&-\frac6{F_1(0)}\frac{dF_1(\mathbf{Q}^2)}{d\mathbf{Q}^2}\Bigg|_{\mathbf{Q}^2\to0}\nonumber\\
&=&-\frac6{F_1(0)}\frac{\partial}{\partial\mathbf{Q}^2}\int\frac{d^3\kappa}{(2\pi)^3}\left[1+\frac{\bm{\kappa}^2-(\mathbf{Q}/2)^2}{4\left\langle\kappa^0\right\rangle^2}\right]\varphi\left(\bm{\kappa}+\frac{\mathbf{Q}}2\right)\varphi\left(\bm{\kappa}-\frac{\mathbf{Q}}2\right)\Bigg|_{\mathbf{Q}\to0}\nonumber\\
&=&a^2\left(\frac32+\frac1{1+\frac{8\left\langle\kappa^0\right\rangle^2a^2}3}\right).
\end{eqnarray}

Another constraint on the model parameters comes from the proton
magnetic moment, which in our additive quark model with charges
$e_u=\frac23e_p$, $e_d=-\frac13e_p$ appears to equal
\begin{equation}\label{mu-p}
    \frac{\mu_p}{e_p}=\frac{\mu_q}{e_q}=\frac1{2\left\langle\kappa^0\right\rangle}F_2(0)=\frac1{2\left\langle\kappa^0\right\rangle}\int\frac{d^3\kappa}{(2\pi)^3}\varphi^2\left(\kappa\right)
=\frac1{2\left\langle\kappa^0\right\rangle}\frac1{1+\frac3{8\left\langle\kappa^0\right\rangle^2a^2}}.
\end{equation}

Together, constraints (\ref{rch-def}-\ref{mu-p}) may determine both
free parameters of the wave function. With the experimental values
${\mu_p}/{e_p}\approx\frac{2.79}{2M_N}$, $\left\langle
r^2_{\mathrm{ch}}\right\rangle\approx(0.9\,\mathrm{fm})^2$
\cite{TW}, Eqs.~(\ref{rch-def}, \ref{mu-p}) are, strictly speaking,
mutually inconsistent (see Fig.~\ref{fig:m-a}). But within our
accuracy $30\div40\%$ they are sufficiently consistent, yielding
$a=0.55\div0.85$ fm and
$\left\langle\kappa^0\right\rangle=50\div250$ MeV. Similar values
for $a$ were found by other authors \cite{Myhrer}.
\footnote{However, if instead of (\ref{ansatz}) one took a model
$\varphi\propto
\left(1+\frac{a^2\bm\kappa^2}{2\alpha}\right)^{-\alpha}$ with
$\alpha<4$, the value for $a$ might be larger.} As for
$\left\langle\kappa^0\right\rangle$, it is about twice smaller than
typical quark mass in non-relativistic constituent models \cite{TW}.

\paragraph{Valence quark distribution function}

The constructed model should also be able to describe valence quark
distribution at Bjorken $x<0.4$, i. e., including the region of the
quark distribution maximum. The definition of a quark parton
distribution (cf., e.g., \cite{Barone}) leads to
\begin{eqnarray}\label{q}
    q_v(x)&=&\int\frac{d^3\kappa}{(2\pi)^3}\delta\left(x-\frac{\kappa^0+\kappa^3}{M_N}\right)\bar\psi_q(\kappa)\left(\gamma^0+\gamma^3\right)\psi_q(\kappa)\nonumber\\
    &=&\frac{M_N}{2\pi}\int\frac{d^2\kappa_\perp}{(2\pi)^2}\left[\left(1-\frac{\kappa^0}{2\left\langle\kappa^0\right\rangle}+\frac{xM_N}{2\left\langle\kappa^0\right\rangle}\right)^2+\frac{\bm{\kappa}^2_\perp}{4\left\langle\kappa^0\right\rangle^2}\right]\varphi^2\left(\kappa\right)\Big|_{\kappa^3=xM_N-\kappa^0}\nonumber\\
    &=&\frac{M_Na}{\sqrt\pi\left(4\left\langle\kappa^0\right\rangle^2a^2+\frac32\right)}\left[\left(xM_N-\kappa^0+2\left\langle\kappa^0\right\rangle\right)^2a^2+1\right]e^{-(xM_N-\kappa^0)^2a^2}.
\end{eqnarray}
\[
\left(\int^\infty_{-\infty}q_v(x)dx=1\right)
\]

\begin{figure}[b!]
  \centering
  \begin{tabular}{cc}
    \includegraphics[width=59mm]{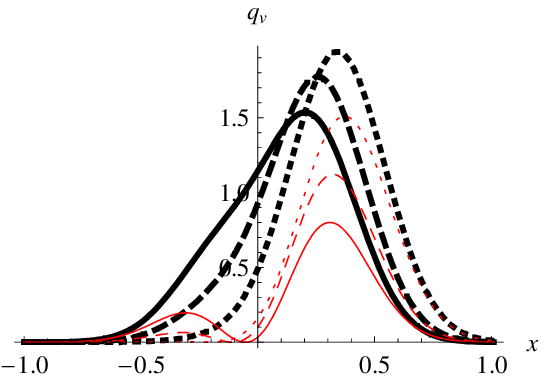} &
    \includegraphics[width=59mm]{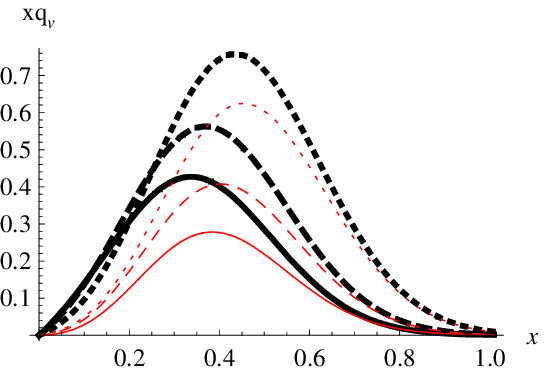}\\
    \textbf{(a)} & \textbf{(b)}
  \end{tabular}
  \caption{%
    \textbf{(a)} Valence quark distribution for $a=0.7$ fm and $m=50$ MeV (solid line),
    $m=100$ MeV (dashed), $m=200$ MeV (dotted). In thin red the transverse spin asymmetry $\triangle_Tq(x)$ is shown.
    \textbf{(b)} Same as (a) but for the valence quark distribution weighted with
    $x$. The momentum sums are, correspondingly,
    $3\int_0^1dxxq_v(x)=$ 0.5, 0.7, 1.0.
  }
  \label{fig:q}
\end{figure}

Behavior of function (\ref{q}) at different
$\left\langle\kappa^0\right\rangle$ is shown in Fig.~\ref{fig:q}.
Values $\left\langle\kappa^0\right\rangle>$150 MeV look rather
unrealistic. At $\left\langle\kappa^0\right\rangle\simeq50$ MeV our
$xq_v(x)$ holds within the declared $30\div40\%$ accuracy with the
MRST, CTEQ pdf fits. It seems interesting that the constituent quark
``mass" $\left\langle\kappa^0\right\rangle=50\div100$ MeV gets
already commensurable with the pion mass scale $m_\pi\approx140$
MeV.

The features of Fig.~\ref{fig:q}a are $\int_0^1dxq_v(x)<1$ and high
value of $q_v(0)$. \footnote{$q_v(0)$ is finite also in bag models
\cite{Jaffe}, whereas in light-cone models often
$q_v(x)\underset{x\to0}\to0$. The latter property seems
phenomenologically unlikely, even assuming further Regge
enhancements.} That is expectable: since the quark rest energy is
smaller than its typical momentum
$\sim(0.7\,\mathrm{fm})^{-1}\simeq300\,\mathrm{MeV}$, the wave
function must reach well across the light cone. Note that $q_v(x)$
at negative values of $x$ should \emph{not} be associated with
antiquarks \cite{Barone}, but still they may describe quarks, only
ones inaccessible to leading twist DIS. In fact, for strongly bound
relativistic states the positivity condition does not apply to
light-cone components of momenta. Even con-sidering the DIS handbag
diagram in terms of 2-particle intermediate states with
Bethe-Salpeter wave functions in nucleon vertices, the
$\kappa^-$-integration will give a non-zero result for $\kappa^+$
beyond the interval $[0,P^+]$, because due the BS vertex functions
$G(\bm\kappa)$ depending on $\kappa^+$ through $\kappa^3$ one can
not completely withdraw the integration contour to $\infty$.

In conclusion, let us point out that the continuity of a valence
quark wave function around $x=0$ permits its use for perturbative
description of hadron-hadron reactions with single quark exchange.
Thereat, non-zero $q_v(0)$ opens a specific possibility of
transverse polarization generation in reaction $np\to pn$
\cite{Bondarenco}.

\end{document}